\documentclass[10pt,traditabstract,A4]{aa}

\usepackage{txfonts}
\usepackage{graphicx}
\usepackage{natbib}

\newcommand{\cotan}{\; {\rm cotan}}

\newcommand{\uin}{u_{\rm in}}

\newcommand{\uout}{u_{\rm out}}
\newcommand{\ain}{a_{\rm in}}
\newcommand{\aout}{a_{\rm out}}

\newcommand{\kint}{k_{\rm in}}
\newcommand{\kout}{k_{\rm out}}

\newcommand{\elik}{{\mathbf K}}
\newcommand{\md}{M_{\rm d}}
\newcommand{\tm}{\tilde{M}}
\newcommand{\psic}{\psi_{\rm c}}
\newcommand{\tpsi}{\tilde{\psi}}

\def\atan{{\rm atan \,}}

\newcommand{\ms}{m_{\rm s}}

\newcommand{\tir}{\tilde{r}}
\newcommand{\tpsis}{\tilde{\psi}_\lambda}
\newcommand{\psis}{\psi_\lambda}
\newcommand{\Ss}{S_\lambda}
\newcommand{\ks}{k_\lambda}
\newcommand{\ksin}{k_{\lambda, \rm in}}
\newcommand{\ksout}{k_{\lambda,\rm out}}
\newcommand{\Lambdas}{\Lambda_\lambda}

\begin{document}

\title{The Newtonian potential of thin disks}
\author{Jean-Marc Hur\'e$^{1,2}$ and Franck Hersant$^{1,2}$}
\institute{Universit\'e de Bordeaux, OASU, 2 rue de l'Observatoire, BP 89, F-33271 Floirac Cedex, France
\and
CNRS, UMR 5804, LAB, 2 rue de l'Observatoire, BP 89, F-33271 Floirac Cedex, France}
\date{Received ??? / Accepted ???}


\abstract{The one-dimensional, ordinary differential equation (ODE) 
that satisfies the midplane gravitational potential of truncated, flat power-law disks is extended to the
whole physical space. It is shown that thickness effects (i.e. non-flatness) can be easily accounted for
by implementing an appropriate ``softening length'' $\lambda$. The solution of this ``softened ODE'' has the following properties: i) it is regular
at the edges (finite radial accelerations), ii) it possesses the correct long-range properties, iii) it matches the Newtonian potential of a geometrically thin disk very
well, and iv) it tends continuously to the flat disk
solution in the limit $\lambda \rightarrow 0$. As illustrated by many examples, the ODE, subject to exact Dirichlet conditions, can be solved numerically with efficiency for any given colatitude at second-order from center to infinity using radial
mapping. This approach is therefore particularly well-suited to generating grids of gravitational forces in order to study particles moving under the field of a gravitating disk as found in various contexts (active nuclei, stellar systems, young stellar objects). Extension to non-power-law surface density profiles is straightforward through superposition.
Grids can be produced upon request.
}


\maketitle

\section{Introduction}

Flattened astrophysical objects like disks are known to genrally produce gravitational fields weaker than spherical bodies of comparable mass. However, the gravity of low-mass disks evolving
on long time scales may play a significant role in their own dynamics and environment
\citep{gt78,subrkaras05}. Gravitational forces from disks are not easily accessible by numerical
computation, and this domain still represents an interesting challenge. For various reasons (misknowledge of
boundary conditions, sensitivity and inaccuracy of solutions, relevant physical scales, kernel singularities,
computational cost, etc.), neither the Poisson equation nor Newton's integral law offers a simple and straightforward tool, and each must be handled with some caution. Truncated expansions of solutions generally suffer from inaccuracy and instability
\citep{clement74,hachisu86}. Softened Gravity for continuous systems may help in some circumstances, but the
influence of the softening length | a free-parameter, classically | is spurious, and it fundamentally destroys
the Newtonian character of the gravitational interaction \citep{he88,ars89}. Each disk configuration (symmetry,
edges, mass profile, shape, etc.) must therefore be investigated individually for a given application.

Geometrically thin disks probably constitute the main class of astrophysical disks. These exhibit various shapes, density profiles and sizes. For those orbiting a massive central object (star or black hole),
a self-similar behavior may develop secularly, leading to a mass density profile varying close to a
power law of the radius. Such a profile is widely supported by theory and it is a typical initial ingredient of
numerical simulations \citep[e.g.][]{pringle81,edgar07}. Even in quasi-Keplerian
rotation, thin disks can be
influenced by their own gravity. \cite{hh07} (hereafter Paper I) have shown that the midplane gravitational
potential of flat, power-law disks obey an ordinary differential equation (ODE) accounting for edges which are
usually ignored \citep{bisnovatyi75,ge99}. Analytical solutions in the form of very rapidly converging
series have been reported in \cite{hhcb08}. 
At the same time, \cite{hp09} have shown that the model of ``Softened Gravity'' offers a good framework for determining the
Newtonian potential of thin disks (whatever the mass density profile), provided the ``softening length''
takes a very specific form,  locally. 
In the present paper, we show that the ODE for the gravitational potential described by \cite{hh07} can be extended to the whole physical
space, and we combine this result with an appropriate softening length to describe the potential of disks of
non vanishing thickness.

The paper is organized as follows. We recall the basic configuration and useful formulae of potential in flat, power-law disks in Sect. \ref{sec:basic}. The derivation of the generalized ODE, non-dimensionalization and asymptotic behavior of solutions, are found in Sect. \ref{sec:ugode}. The numerical solutions are given in Sect. \ref{sec:num}. Thickness effects, including the introduction of the softening length, are discussed in Sect \ref{sec:lambda}. The last section is devoted to a conclusion.

\section{Theoretical grounds and notation}
\label{sec:basic}

\begin{figure}[h]
\centering
\includegraphics[width=8.9cm]{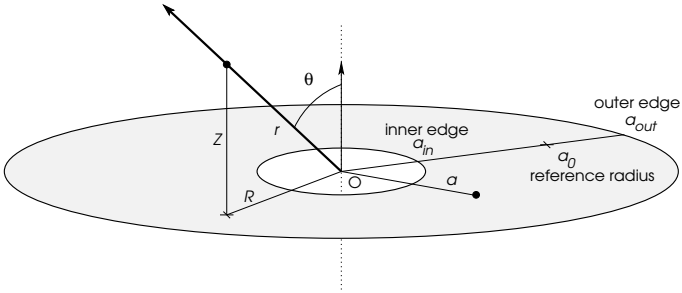} 
\caption{Configuration for the flat, finite size disk with radius at the edges $\ain$ and $\aout$; $r$ and $\theta$ are spherical coordinates, $R$ and $Z$ are cylindrical coordinates.}
\label{fig:scheme2.xfig}
\end{figure}

Following \cite{hh07} (hereafter Paper I), we consider a flat axisymmetrical disk with inner edge $\ain \ge 0$, outer edge $\aout > \ain$ (see Fig. \ref{fig:scheme2.xfig}), and a power-law surface density of the form
\begin{equation}
\Sigma(a) = \Sigma_0 \left(\frac{a}{a_0}\right)^s,
\end{equation}
where $a$ is the cylindrical radius, $a_0$ some reference radius, and $\Sigma_0 \equiv \Sigma(a_0)$ the corresponding surface density. This profile can serve as a basis for defining more complex mass distributions, by mixing power laws with different indices (positive and negative). For such a disk, the gravitational potential in space is given exactly by the expression \citep[][]{durand64}
\begin{equation}
\psi(\vec{r}) = -2G \int_{\ain}^{\aout}{\sqrt{\frac{a}{R}} \Sigma(a) k \elik(k)da},
\label{eq:psi}
\end{equation}
where 
\begin{equation}\elik(k)=\int_0^{\pi/2}{\frac{d\phi}{\sqrt{1-k^2 \sin^2 \phi}}}
\label{eq:elik}
\end{equation}
is the complete elliptic integral of the first kind, and
\begin{equation}
k=\frac{2\sqrt{aR}}{\sqrt{(a+R)^2+Z^2}}
\label{eq:kmodulus}
\end{equation}
 is the modulus ($0 \le k \le 1$), $R$ and $Z$ are the cylindrical coordinates, and $r$ is the spherical radius (i.e. $r^2 =R^2+Z^2$). Known properties about this configuration are the followings. The integral in Eq. \ref{eq:psi} has a diverging kernel as soon as the modulus $k$ reaches unity. This occurs everywhere inside the disk. Standard quadrature schemes fail to give accurate potential values unless a specific treatment is considered \citep{hurepierens05}. The potential is not a power law of the radius, because of edges. A closed form for Eq. \ref{eq:psi} exists only in the case of infinitely extended disks \citep{bisnovatyi75,ge99} and for finite size disks with constant surface density \citep{durand64,lassblitzer83}. The potential is finite everywhere, except when $\ain=0$ and $s+1<0$, but has an infinite radial gradient in the midplane at edge crossing, as a consequence of flatness \citep[][]{durand64,mestel63}. Finally, the disk mass is finite except when $\ain=0$ and $s+2<0$, or when $\aout \rightarrow \infty$ and $s+2>0$.

It has been established in Paper I that the potential $\psi$ given by Eq. \ref{eq:psi} obeys, {\it in the disk midplane} (i.e. for $Z=0$), a differential equation of the form
\begin{equation}
\frac{d \psi}{d R} - (1+s) \frac{\psi}{R} - \Lambda=0,
\label{eq:ode}
\end{equation}
where $\Lambda$ is fully analytical. Solutions in the form of rapidly converging series were reported in \cite{hhcb08}.

\section{Generalized ODE}
\label{sec:ugode}

The starting point of the generalization of Eq. \ref{eq:ode} is the equation of the line containing the origin and making an angle $\theta  \in [0,\pi]$ (the colatitude) with the $z$-axis (see Fig. \ref{fig:scheme2.xfig}). This equation is simply
\begin{equation}
Z= R \cotan \, \theta.
\label{eq:zr}
\end{equation}
By setting $u=a/R$ in  Eq. \ref{eq:kmodulus}, and using Eq. \ref{eq:zr}, we get the relation
\begin{equation}
u^2 - 2u \left(\frac{2}{k^2}-1\right)+1+\cotan^2 \theta=0.
\label{eq:uofk}
\end{equation}
We then see that, once $\theta$ is specified, $u$ is a function of the modulus $k$ only. This property was the
condition for the existence of Eq. \ref{eq:ode}. It still holds here. We can therefore proceed as in Paper I : the partial derivative of the potential with respect to the radius can be expressed as a function of the potential itself. Here, we consider the derivative with respect to the spherical radius $r$ instead of the cylindrical radius $R$ (although this choice is not really important). The full demonstration is given in Appendix \ref{sec:ap:derivation}. Strictly speaking, we obtain a partial differential equation (PDE) that becomes, {\it for a given colatitude $\theta$}, an ordinary differential equation (ODE). It writes as
\begin{equation}
\frac{d \psi}{d r} - (1+s) \frac{\psi}{r} - \Lambda=0,
\label{eq:gode}
\end{equation}
where $\Lambda$ is defined by
\begin{equation}
\Lambda = 2 G \Sigma_0 \frac{a_0}{r} u_0^{-(s+1)} \left[\uout^{s+\frac{3}{2}} \kout \elik(\kout)  - \uin^{s+\frac{3}{2}} \kint  \elik(\kint) \right],
\end{equation}
with $\uout=\aout/R$, and $\uin=\ain/R$. This generalizes to the whole space the differential equation reported  in Paper I, which is valid only in the disk plane (i.e. the case $\theta=\frac{\pi}{2}$). The second member $\Lambda$ is an analytical function of a single spatial coordinate $r$. It now depends on the colatitude $\theta$, and on the four parameters $s$, $a_0$, $\ain$, and $\aout$. The main difference comes from the $Z$-dependent modulus $k$.

We can make the ODE scale-free by setting\footnote{The choice $a_0=\ain$ (or $a_0=\aout$) is possible but not appropriate in the particular case where $\ain=0$ (respect. $\aout \rightarrow \infty$).} $\tir = r/a_0$, and $\tpsi = \psi/\psi_0$ where $\psi_0$ is a constant. Then, we get the non-dimensional ODE:
\begin{equation}
\frac{d \tpsi}{d \tir} - (1+s) \frac{\tpsi}{\tir} = S,
\label{eq:gode2}
\end{equation}
where
\begin{equation}
 S = \Lambda \frac{a_0}{\psi_0}.
\end{equation}

Potential values are known at the two edges of the disk in the form of rapidly converging series \citep{hhcb08}, and can therefore be used for adimensioning. A more convenient value for $\psi_0$ is probably the potential at the origin which is exact. From Eq. \ref{eq:psi}, we have
\begin{eqnarray}
\nonumber
\psi(\vec{0}) & = -2 \pi G \Sigma_0 a_0 u_0^{-(s+1)}\int_{\ain}^{\aout}{u^s du}\\
& = - 2 \pi G \Sigma_0 a_0 u_0^{-(s+1)} \uout^{s+1} \chi_{s+1} \equiv \psic,
\label{psiatthecenter}
\end{eqnarray}
where we have set (definition different than in Paper I)
\begin{equation}
\chi_n = \frac{1 - \Delta^{n}}{n},
\end{equation}
and $\Delta=\ain/\aout$ is the axis ratio of the disk. Thus, with $\psi_0 \equiv \psic$, the inhomogeneous term in the non-dimensional ODE is
\begin{equation}
S = - \frac{1}{\pi \chi_{s+1} \tir} \left[ \sqrt{\uout} \kout \elik(\kout)  - \Delta^{s+1}  \sqrt{\uin} \kint  \elik(\kint) \right].
\label{eq:sinhomo}
\end{equation}
When $\ain=0$ (no inner edge; the disk extends down to the center), the last term vanishes. When $\aout \rightarrow\infty$, this is the first term. The asymptotic properties of the solution $\tpsi$ are discussed in Appendix \ref{sec:ap:asymptotics}.

\begin{figure}
\includegraphics[width=8.9cm,bb=3 2 786 544,clip=]{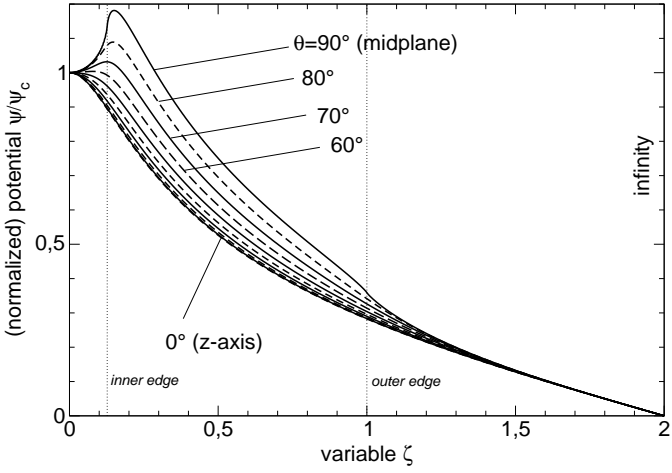} 
\caption{Potential versus $\zeta$ due to a flat disk with $\Delta = 0.1$, $\aout/a_0 = 1$ and power index $s=-1.5$,
for several colatitudes $\theta$. Values are normalized to the central value $\psic$. The grid is such that
$A=\pi/4$ and $N=1000$. Infinity stands at $\zeta=2$, and the edges are located at $\zeta \approx 0.1269$ and
$1$ (for $\theta=\frac{\pi}{2}$). Although not visible on the figure, the potential is not derivable at both
edges (case with $\theta=\frac{\pi}{2}$; see Sect. \ref{sec:basic}).} 
\label{fig:pot-1.5.eps} 
\end{figure}      

\section{Numerical solution in space with mapping}
\label{sec:num}

There are many methods for solving a first-order ODE numerically. This needs one boundary condition for an initial value
problem (IVP), but two conditions for a two boundary value problem (TBVP). The potential is known at four places: at the
origin, at the two edges, and at infinity. The origin is particularly well suited to finding the potential in
space in a given direction $\vec{r}$, as the corresponding boundary condition is the same regardless of the
colatitude. Values at the disk edges are useful mainly for determining the potential inside the disk or in its neighborhood. For some applications, potential values are required at large distances from the disk \citep[e.g.][]{skh04,subrkaras05}. In this case, it can be interesting to use the trivial boundary condition at infinity ($\psi \rightarrow 0$ as $r \rightarrow \infty$). This is easily performed by mapping the whole space. For instance, the transformation\footnote{\label{note:tang}Since the tangent function is essentially linear with the argument in the range $[0,\pi/4]$, the choice $A = \pi/4$ leaves the space inside the shell $r \le a_0$ almost undistorted by this change of coordinate.}
\begin{equation}
\zeta(\tir) = \frac{1}{A}\atan \tir,
\end{equation}
where $A$ is a positive constant, maps the range $\tir \in [0,\infty[$ into the compact domain $\zeta\in [0,\frac{\pi}{2A}]$. The ``mapped ODE'' reads
\begin{equation}
 \frac{d \tpsi}{d \zeta}- (1+s) f \frac{\tilde{\psi}}{\zeta}  - f \frac{Q}{\zeta} =0,
\label{eq:tbvpmapping}
\end{equation}
where we have set
\begin{equation}
\frac{1}{f} \equiv  \mathrm{sinc}(2 A \zeta),
\end{equation}
and
\begin{equation}
Q = S \tir.
\end{equation}

It is then easy to discretize Eq. \ref{eq:tbvpmapping} on a grid, regular in $\zeta$, and to solve the associated linear system by standard techniques. Actually, a centered second-order space discretization based on $N+1$ mesh points yields the system:
\begin{equation}
\left\{
\begin{array}{l}
\label{eq:psigen2}
\nonumber
\tpsi_0=1,\\\\
\tpsi_{n-1} + 2 \frac{\delta \zeta}{\zeta_n}f_n (1+s) \tpsi_n - \tpsi_{n+1} + 2 \frac{\delta \zeta}{\zeta_n} f_n Q_n =0,\\
\qquad n \in \{1,\dots,N-1\},\\\\
\nonumber
\tpsi_N=0.
\end{array}
\right.
\end{equation}
with $\zeta_0=0$, $\zeta_N = \frac{\pi}{2A}$, and $\zeta_n=n \delta \zeta$. For $s=-1$, it can be advantageous to rewrite the ODE in a slightly different form, for instance by using the auxiliary function $\tm \equiv \tpsi \tir$, which satisfies the following ODE:
\begin{equation}
 \frac{d \tm}{d \zeta}- (2+s)f \frac{\tm}{\zeta} - \frac{1}{\cos^2(A\zeta)}Q=0,
\label{eq:tbvpmapping-m}
\end{equation}
with the boundary conditions $\tm(0)=0$ and $\tm\left(\frac{\pi}{2A}\right)=\frac{\Delta-1}{\ln \Delta}$.

\begin{figure}[h]
\includegraphics[width=8.9cm,bb=3 2 786 544,clip=]{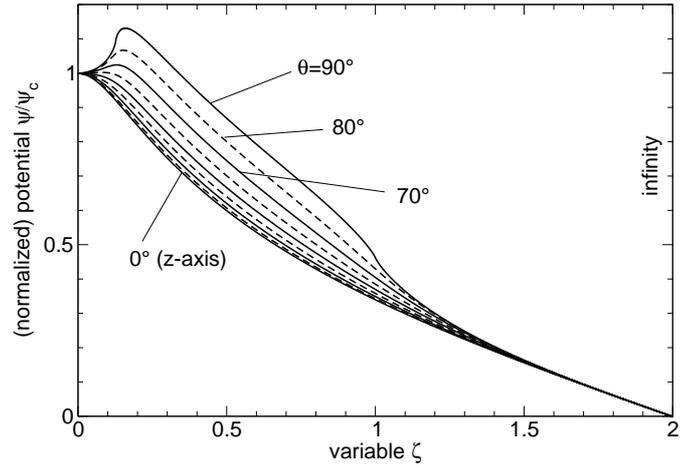} 
\caption{Same legend as for Fig. \ref{fig:pot-1.5.eps}  but for $s=-1$. In this case, we used the auxiliary function $\tm$ and Eq \ref{eq:tbvpmapping-m}.} 
\label{fig:pot-1.eps} 
\end{figure}

\begin{figure}[h]
\includegraphics[width=8.9cm,bb=3 2 786 544,clip=]{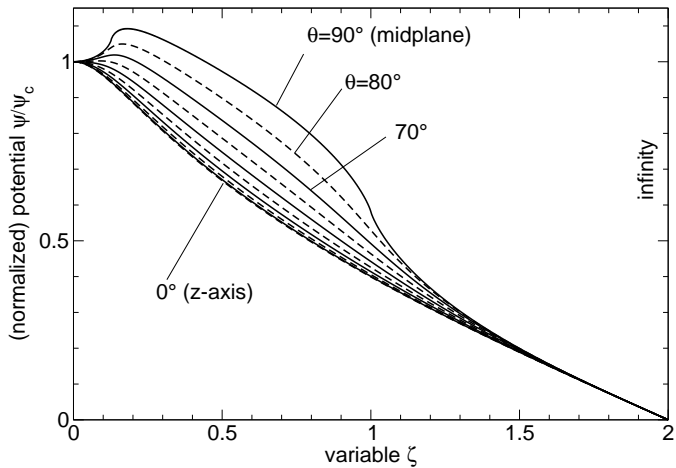} 
\caption{Same legend as for Fig. \ref{fig:pot-1.5.eps}  but for $s=-0.5$.} 
\label{fig:pot-0.5.eps} 
\end{figure}

\begin{figure}[h]
\includegraphics[width=8.9cm,bb=3 2 786 544,clip=]{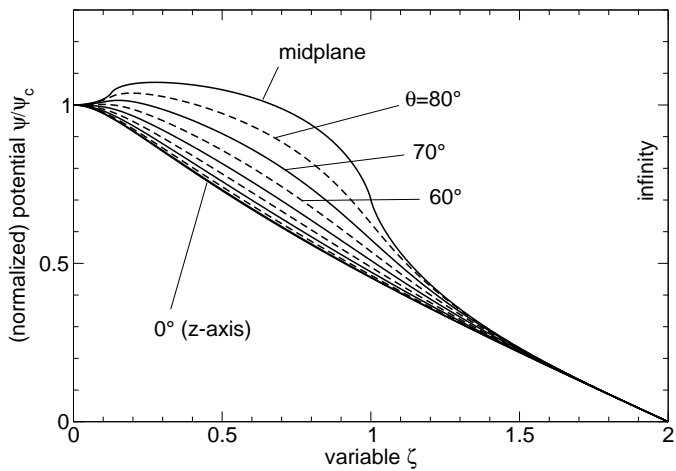} 
\caption{Same legend as for Fig. \ref{fig:pot-1.5.eps}  but for $s=0$ corresponding to an homogeneous disk.} 
\label{fig:pot-0.eps} 
\end{figure}

 To illustrate this numerical part, we show in figure \ref{fig:pot-1.5.eps} the solution $\{(\zeta_i,\tilde{\psi}_i)\}$ obtained at a few colatitudes with $N=1000$ and $A=\frac{\pi}{4}$ (see note \ref{note:tang}) and for the following parameters: $\Delta = 0.1$, $\aout/a_0 = 1$, and $s=-1.5$ (due to adimensioning, it is not necessary to specify $\Sigma_0$). Figures  \ref{fig:pot-1.eps} to \ref{fig:pot-0.eps} correspond to power indices $s=\{-1,-0.5,0\}$.

The time required to solve the above system is linear with $N$. A full grid is then generated in a time that is proportional to $N \times M$, where $M$ is the number of mesh points in the $\theta$-direction. With a classical laptop, this takes about $15$s to generate a grid with $N^2=10^6$ mesh points (including the update of the inhomogeneous term $Q$). This is much faster than what can be obtained from the multipole expansion, by a factor proportional to the number of Legendre polynomials to be considered, which is generally very large | typically, several hundreds inside the source \cite[][]{clement74}.

\begin{figure}[h]
\includegraphics[width=7.6cm,bb=47 145 452 592,clip=,angle=-90]{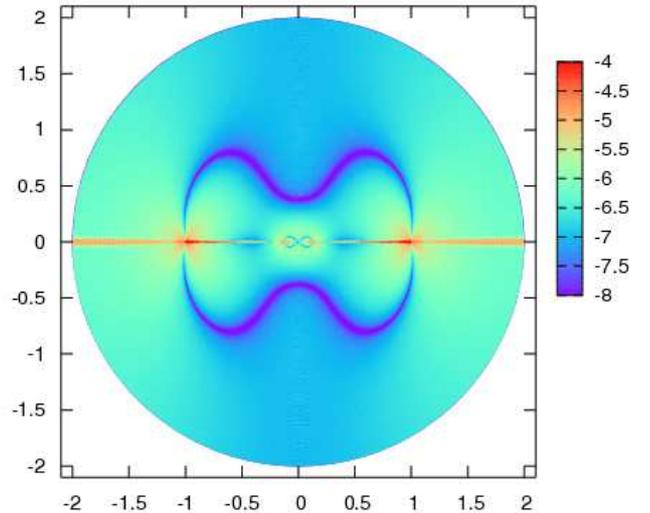} 
\caption{Logarithm of the error relative to exact values for the homogeneous case (same conditions as for Fig. \ref{fig:pot-0.eps}). The grid has $N=1000$ and $M=100$ (signed $\zeta$-variable).}
\label{fig:pot_map0.eps}
\end{figure}   

Accuracy can be properly checked in the homogeneous case since the potential is known in a closed form
\citep{durand64,lassblitzer83}. Figure \ref{fig:pot_map0.eps} displays the error relative to the exact value. The radial grid and the disk parameters are the same as for Fig. \ref{fig:pot-0.eps} (we used $M=100$ for the sampling in $\theta$). We see that the agreement is very good in
the whole space, less than $10^{-6}$ on average for this radial resolution, and it is rather uniform in the
whole space (however, with a certain deterioration in the midplane due to edges; see Sect. \ref{sec:basic}).
The accuracy only depends on the discretization of the ODE, that is, only on $N$ (not on $M$).
Increasing $N$ obviously improves potential values.

\section{Thickness effects and softening length}
\label{sec:lambda}

The flat disk hypothesis is well-suited to many studies, both theoretical and numerical ones. Its physical
realism is, however, limited. For instance, gas or particles present at each edge would feel an infinite acceleration there (see again Sect. \ref{sec:basic}). Such singularity would disappear by considering an extra dimension (or with a vanishing density profile at the disk edges).  Thickness effects (matter present off the midplane) can therefore be important, not only for thermodynamic reasons, but also from a dynamical point of view. We have not yet investigated the existence of an ODE corresponding to geometrically thin (i.e. non-flat) disks. However, we can reproduce {\it the Newtonian potential due to a thin disk with a one dimensional differential equation} of the form of Eq. \ref{eq:gode} by a ``softened'' potential\footnote{For the concept of Softened Gravity in point mass systems, see \cite{he88}, and see \cite{ars89} for the case of gas disks.}. In this context, \cite{hp09} showed that the midplane potential of a vertically stratified disk can be approximated by an equation resembling Eq. \ref{eq:psi} provided the modulus $k$ is changed for the ``softened modulus'':
\begin{equation}
\ks = \frac{2\sqrt{a R}}{\sqrt{a^2+ r^2 +2 aR + \lambda^2}},
\label{eq:klambda}
\end{equation}
where $\lambda$ is called the ``softening length'' (generally considered as a free parameter). They found that the best expression for $\lambda$ that preserves the Newtonian character of the disk potential takes the form
\begin{equation}
\lambda \approx h\times g(a,R,\dots),
\label{eq:lambda}
\end{equation}
where $h$ is the local semi-thickness of the disk, and $g$ is a {\it bounded, slowing varying function of the
radius}. This result is limited to disks with small aspect ratios, locally, i.e. to geometrically thin disks in the classical sense. This result has confirmed the common idea that $\lambda$ is a certain fraction of the disk
thickness. In the absence of mass density gradients in the direction perpendicular to the disk equatorial
plane, $g$ goes through a sharp minimum of $e^{-1} \approx 0.368$ at $R=a$ (i.e. for $u=1$). The sensitivity with stratification appears weak since for a vertical parabolic profile, this minimum is
$e^{-4/3} \approx 0.264$ and the sharpness of the peak is unchanged. Appendix \ref{sec:ap:sl_asymptotics}
reproduces the result preliminarily derived in the limit of $g$ far from
kernel singularity (i.e. for $u\rightarrow 0$ or $\infty$). In particular, we find that $g$ tends
asymptotically to $3^{-1/2} \approx 0.577$ in the homogeneous case, while it is $5^{-1/2} \approx 0.447$ in
the parabolic case. We conclude that, for a given vertical profile, the function $g$ does not vary much,
except in the vicinity of $R=a$ where it decreases by a factor of $ \sim 37\%$. As $\lambda$ does not
depend on the precise shape $h(a)$ provided the disk has a small aspect ratio \citep[see][]{hp09}, this conclusion seems quite robust. This point is confirmed by the numerical experiments. In the following, we therefore neglect any radial variation of $g$, and consider that {\it it is a constant} whose nominal value will be selected a posteriori (and this assumption will be justified). Now, reversing Eq. \ref{eq:klambda} as done before and using Eq. \ref{eq:lambda}, we find at a given colatitude
\begin{equation}
u^2 \left(1 +  g^2 \epsilon^2 \right) - 2u \left(\frac{2}{\ks^2}-1\right)+1+\cotan^2 \theta =0,
\label{eq:uofk_withlambda}
\end{equation}
where we have set $\epsilon=h/a$. It means that, provided the {\it aspect ratio of the disk $\epsilon$ is a constant}, the ODE can still be formed. After some algebra and with the same non-dimensionalization as considered above, we actually find
\begin{equation}
\frac{d \tpsis}{d \tir} - (1+s) \frac{\tpsis}{\tir} - \Ss=0,
\label{eq:odelambda}
\end{equation}
where:
\begin{eqnarray}
\label{eq:slambdainhomo}
\Ss  = - \frac{1}{\pi \chi_{s+1} \tir} & \left[ \sqrt{\uout} \ksout \elik(\ksout)\right.\\  &\left. - \Delta^{s+1}  \sqrt{\uin} \ksin  \elik(\ksin) \right],
\nonumber
\end{eqnarray}
where $\ksin \equiv \ks$ at $a=\ain$, and $\ksout \equiv \ks$ at $a=\aout$. We see that $\Ss$ only differs from Eq.
\ref{eq:sinhomo} by the presence of softened modulus $\ks$ (instead of $k$). Since we have
\begin{equation}
\ks \le \frac{1}{\sqrt{1+\epsilon^2g^2}}<1,
\end{equation}
it follows that $\Ss$ is no longer singular at the disk edges. As a result, the radial acceleration at the edges becomes finite. The role of $\lambda$ is to mimic an extra dimension (the vertical one). 

The softened potential $\tpsis$ entering Eq. \ref{eq:odelambda} can then be solved numerically in the whole
space, as done for the flat disk. In this process, the inner boundary value is slightly lower and must be changed to
\begin{equation}
\tpsi_{\rm s,0} = \frac{\psis (\vec{0})}{\psic} = \frac{1}{\sqrt{1+\epsilon^2g^2}} < 1.
\end{equation}

By looking at Eqs. \ref{eq:gode} and \ref{eq:odelambda}, we see that there is a perfect continuity between these two ODEs (with and without softening length) as  $\lambda \rightarrow 0$. It means that $\tpsis \rightarrow \tpsi$  as $\lambda \rightarrow 0$ (including the value at the origin). Also, it is worth noting that the {\it explicit dependence with $\lambda$ must be regarded as a dependence with the disk thickness}.

\begin{figure}[h]
\includegraphics[width=8.9cm,bb=28 8 720 583,clip=,angle=0]{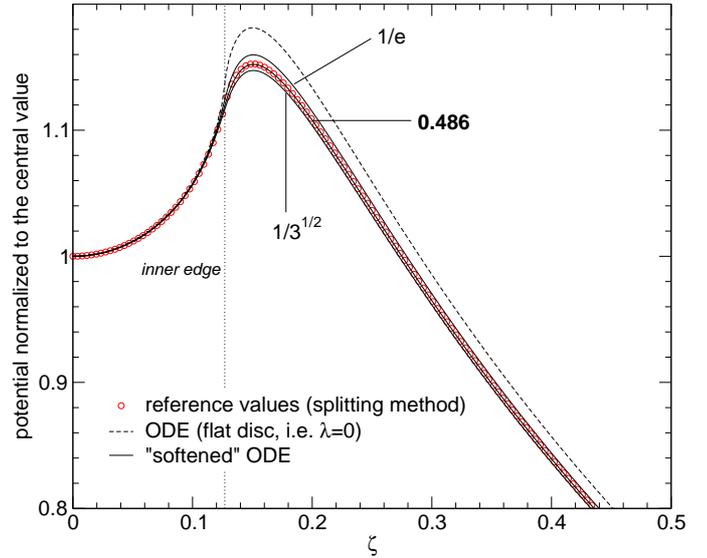} 
\caption{Comparison of different potentials: Newtonian potential of a geometrically disk with $\epsilon=0.1$ ({\it red circles}), Newtonian potential of a flat disk ({\it dashed line}), and softened potential with $g = \{\frac{1}{e}, 0.486, \frac{1}{\sqrt{3}} \}$ ({\it thin lines}). Values are normalized to their central value.  In all cases, $\Delta=0.1$, $\aout/a_0=1$, and $s=-1.5$ (and the same disk mass).}
\label{fig:potwithlambda.eps}
\end{figure}

\begin{figure}[h]
\includegraphics[width=8.9cm,bb=39 8 711 591,clip=,angle=0]{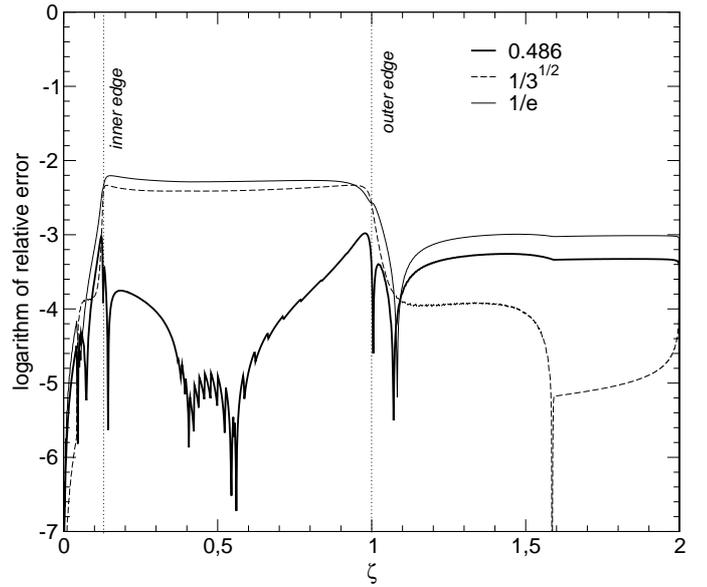} 
\caption{Relative deviation between the softened potential determined in the midplane from Eq.
\ref{eq:odelambda} with $g = \{\frac{1}{e}, 0.486, \frac{1}{\sqrt{3}} \}$ and the Newtonian potential of a
geometrically thin disk with same edges and same mass (see also Fig. \ref{fig:potwithlambda.eps}).}
\label{fig:potwithlambda_error.eps}
\end{figure}

\begin{figure}[h]
\includegraphics[width=7.6cm,bb=47 145 452 592,clip=,angle=-90]{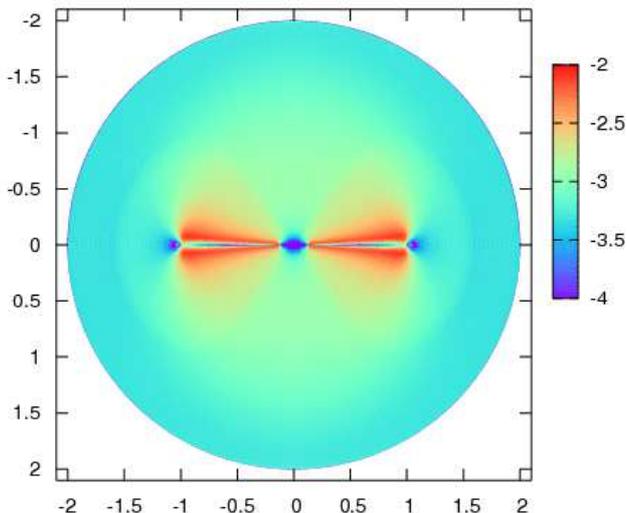} 
\caption{Error map. Same as for Fig. \ref{fig:potwithlambda_error.eps} but at all colatitudes.}
\label{fig:pot_cmp.eps}
\end{figure}

To check the reliability of this approach, we first compared the softened potential $\tpsis$ determined numerically from Eq. \ref{eq:odelambda} with the Newtonian potential of a geometrically thin disk with the same edges, same surface density (and mass), and semi-thickness $h \propto a$ (precisely $\epsilon=0.1$) (hereafter thin disk configuration A). This reference is computed from the splitting method \citep{hure05}, which is very accurate. A typical result, limited to the vicinity of the inner edge, is shown in Fig. \ref{fig:potwithlambda.eps}. We see that the agreement between the two curves is very good. There is a very weak sensitivity to the $g$-parameter. The best agreement is obtained for $g \approx 0.486$, which is very close to the average of bounds. Interestingly enough, thickness effects mainly shift the potential curve, leaving the radial gradients almost unchanged (except at the edges). Figure \ref{fig:potwithlambda_error.eps} shows the logarithm of the relative error in the entire midplane. The error is low, less than $0.1 \%$ inside the disk. With $g=\frac{1}{\sqrt{3}}$, the agreement is even better at a large distance from the disk (but not inside it). Figure \ref{fig:pot_cmp.eps} displays the error for all radii and all colatitudes. We conclude that the global solution of the softened ODE is very close to the Newtonian potential of a thin disk. The largest deviations are observed in the neighborhood of the disk surface (i.e. $Z \approx h$). This is expected since the softened modulus has been approximated here (but this point can be revisited if necessary).

\begin{figure}[h]
\includegraphics[width=8.9cm,bb=13 9 573 805,clip=,angle=0]{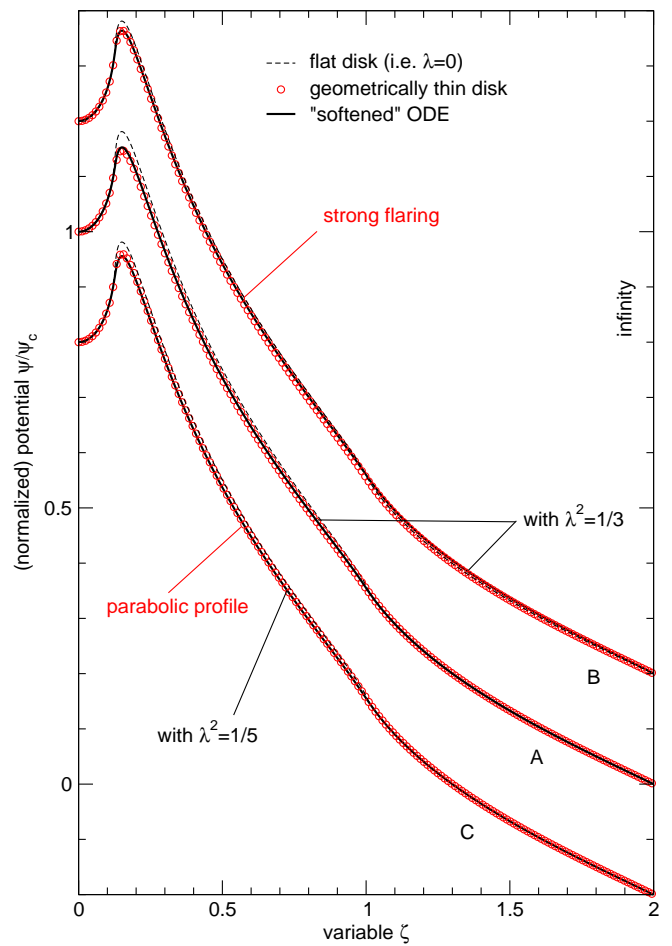} 
\caption{Same legend as for Fig. \ref{fig:potwithlambda_error.eps} but for configuration A, B, and C. For
clarity, potential values are shifted upward for configuration B, and downward for configuration C.}
\label{fig:potwithlambda_4.eps}
\end{figure}

Up to now, only disks with a constant aspect ratio and a homogeneous vertical mass density profile have been considered. The generality of the method can be studied easily by considering other disk configurations, in particular, configurations for which the conditions for the derivation of the ODE are violated. We illustrate this point by considering two other thin disk configurations:
\begin{itemize}
\item configuration B: same as configuration A but the disk is strongly flared, with $h/a = \epsilon (a/\aout)^{0.25}$ (this is an extreme case if we refer to most disk models). In this case, the expression for the softening length is still valid, but the conditions for deriving the softened ODE are violated since $\lambda/h$ is not a constant.
\item configuration C: same as configuration A but the mass density profile is parabolic, vertically. In this case, the
ODE is fully valid, but $g$ must be set to the appropriate value (in between $e^{-4/2}$ and $5^{-1/2}$; see above).
\end{itemize}
In all cases, the disk is geometrically thin at the outer edge (with $\epsilon=0.1$), which is required to keep the softening length valid). We  performed the same comparisons as for configuration A and the results are summarized in Fig. \ref{fig:potwithlambda_4.eps}. We find that the relative deviation between the Newtonian potential of the thin disk and the softened potential (found from the softened ODE) typically varies as shown in Fig. \ref{fig:potwithlambda_error.eps}, and is much less than $1\%$ inside the disk. This is also the case outside the disk, except for configuration B for which the relative error gradually rises from $R=\aout$ and reaches $10\%$ at infinity. We therefore conclude that using the softened ODE to mimic the Newtonial potential of a thin disk with a precision of a few percent is fully justified as soon as the disk is geometrically thin at all radii. The best values are reproduced when the flaring angle of the thin disk is close to zero.

\section{Concluding remarks}

In this article, we have shown that the gravitational potential of a flat, power-law disk can be numerically
determined in the whole physical space by solving, for any colatitude, a linear system resulting from the
second-order discretization of a first-order ordinary differential equation (ODE). The computing time is
especially low because it is proportional to the number of mesh points. It is also demonstrated that the softened
potential based on the prescription for the softening length by \cite{hp09} obeys a similar ODE whose
solution agrees very well with the potential of a geometrically thin disk with same size, mass, and edges and
a constant aspect ratio. We briefly tackle the limits of the method, and conclude that the Newtonian potential is reproduced in the full space, provided i) the disk is geometrically thin {\it at all radii}, which justifies the use of $\lambda$ proposed by \cite{hp09}, and ii) the local flaring remains moderate (i.e. $h/a$ does not vary much with the radius), hence supporting the derivation of the ODE. The implementation of a softening length not only mimics a vertical stratification,
but also totally removes edge singularities typical of flat disks with sharp edges. If the surface density
does not obey a power law but can be written in the form of a series of power laws (e.g. Taylor expansion),
the potential is then obtained easily by superposition (see Paper I). The computing time is then proportional to the number of terms in the series. This approach is especially efficient at generating grids of forces.
This offers reliable tools for investigating the dynamics of particles in a system containing a perturbing, gravitating thin disk \cite[e.g.][]{subrkaras05} and possibly for linking back various families of trajectories to the mass distribution in the disk. The two components of the gravitational force are found from potential values by finite differences. The radial component can, however, be found directly from the ODE. For a unit mass, this component is simply
 \begin{equation}
F_r = - (1+s) \frac{\psis}{r} - \Lambdas.
\end{equation}
It would be interesting to investigate the existence of a first-order differential equation associated with
the colatitudinal gradient of the potential, complementary to the radial ODE reported here. This is the
subject of ongoing work.

\begin{acknowledgements}
It is a pleasure to thank Masters students D. Bernard and J. Lambert. The referee is acknowledged.
\end{acknowledgements}

\bibliographystyle{aa}
\bibliography{../hurebibtex}

\newpage

\appendix

\section{Derivation of the generalized ODE}
\label{sec:ap:derivation}

Starting from Eq. \ref{eq:uofk}, and assuming $\theta$ constant, the derivative of $u$ with respect to $k$ is
\begin{equation}
\frac{du}{dk} = \frac{8u^2 \sin^2 \theta}{k^3 \left(1 - u^2 \sin^2 \theta \right)} \equiv \left.\frac{\partial u}{\partial k}\right|_\theta.
\label{eq:duoverdk}
\end{equation}
This derivative is therefore a function of only $k$. The potential in Eq. \ref{eq:psi} can then be written as the product of a function of the spherical radius $r$ by an integral over $k$, namely
\begin{equation}
\psi(\vec{r}) = -2 G \Sigma_0 a_0^{-s} R^{s+1} \times \int_{\kint}^{\kout}{ {\cal H}(k) dk},
\label{eq:psigen}
\end{equation}
where
\begin{equation}
{\cal H}(k) = u^{s+\frac{1}{2}}  \elik(k)k \times \frac{du}{dk}.
\end{equation}
The new integral bounds, $\kint$ and $\kout$, correspond to the disk edges $\ain$ and $\aout$, and they are found from Eq. \ref{eq:kmodulus}. In particular, we have
\begin{equation}
\kint^2 = \frac{4\ain R}{\ain^2+ r^2 +2 \ain R}
\end{equation}
and
\begin{equation}
\kout^2 = \frac{4\aout R}{\aout^2+ r^2 +2  \aout R},
\end{equation}
respectively. For constant colatitude $\theta$, the derivative of $\psi$ with respect to the spherical radius is then given by
\begin{eqnarray}
\label{eq:psigen2}
\frac{d \psi}{d r} &= (1+s) \frac{\psi}{r} -2G \Sigma_0 a_0^{-s} R^{s+1} \frac{d}{dr} \int_{\kint}^{\kout}{{\cal H}(k) dk}\\
 &\equiv \left. \frac{\partial \psi}{\partial r}\right|_\theta
\nonumber
\end{eqnarray}
This quantity is just the opposite of the radial acceleration due to the disk. Using an elementary derivation rule (see Paper I), we can calculate the right hand-side of this equation.
 We find
\begin{eqnarray}
\frac{d \psi}{d r} = (1+s) \frac{\psi}{r} & \\
&- 2G \Sigma_0 a_0 u_0^{-(s+1)} \left[ {\cal H}(\kout) \frac{d\kout}{dr}  - {\cal H}(\kint) \frac{d\kint}{dr} \right],
\nonumber
\end{eqnarray}
where
\begin{equation}
\frac{d\kout}{dr} \equiv \left.\frac{\partial k}{\partial r}\right|_{\aout},
\end{equation}
and similarly for $\kint$. When $a$ is held constant, we have
\begin{equation}
\left.\frac{\partial k}{\partial r}\right|_a = \frac{k^3 \left( u^2 \sin^2 \theta -1  \right)}{8 a \sin \theta},
\end{equation}
and so
\begin{equation}
\left.\frac{\partial k}{\partial r}\right|_a \times \left.\frac{\partial u}{\partial k}\right|_a = - \frac{a}{r^2  \sin \theta} = - \frac{u}{r}.
\end{equation}
It follows that
\begin{equation}
\frac{d \psi}{d r} - (1+s) \frac{\psi}{r} - \Lambda=0,
\end{equation}
where $\Lambda$ is defined by:
\begin{equation}
\Lambda = 2 G \Sigma_0 \frac{a_0}{r} u_0^{-(s+1)} \left[\uout^{s+\frac{3}{2}} \kout \elik(\kout)  - \uin^{s+\frac{3}{2}} \kint  \elik(\kint) \right],
\end{equation}
$\uout=\aout/R$ and $\uin=\ain/R$. This ordinary differential equation (ODE) is the generalization to the whole space of the ODE reported in \cite{hh07}, which was valid only in the disk plane (i.e. the case $\theta=\frac{\pi}{2}$). This expression differs mainly by the presence of the spherical radius $r$ (instead of $R$) and by the presence of the $Z$-dependent modulus $k$ (instead of $m$). As in Paper I, the second member $\Lambda$ is an analytical function of a single spatial coordinate $r$. It now depends on the colatitude $\theta$ and on the four parameters $s$, $a_0$, $\ain$, and $\aout$.

\section{Asymptotic properties}
\label{sec:ap:asymptotics}

The differential equation possesses the right properties both at small and at large distances. At a large distance from the disk, the modulus in the complete elliptic integrals tends to zero. We have
\begin{equation}
\lim_{k \rightarrow 0} \elik(k) = \frac{\pi}{2},
\end{equation}
and then
\begin{equation}
\lim_{\tir \rightarrow \infty} S = - \frac{(2+s)}{\tir^2}\frac{\chi_{2+s}}{\chi_{1+s}}.
\end{equation}
Since the total mass of the disk is given by the expression
\begin{equation}
\md = 2 \pi \int_{\ain}^{\aout}{\Sigma(a) a da} = - \frac{\psic \aout}{G} \frac{\chi_{2+s}}{\chi_{1+s}},
\label{eq:tmass}
\end{equation}
the above ODE can then be rearranged into
\begin{eqnarray}
\nonumber
r \frac{d \psi}{d r} & \approx (1+s)\psi + (2+s) \frac{G\md}{r}\\ &\approx s \left( \psi + \frac{G\md}{r} \right) + \psi + 2\frac{G\md}{r}
\end{eqnarray}
As this equation must be satisfied for any $s$, we must have
\begin{equation}
\lim_{\tir \rightarrow \infty} \psi = - \frac{G \md}{r},
\end{equation}
which is the expected behavior (the disk is no longer distinguishable from a point mass). This also implies that
\begin{equation}
\lim_{\tir \rightarrow \infty} -\frac{d \psi}{d r} = - \frac{G\md}{r^2}.
\end{equation}

At a short distance around the origin (i.e. $r \ll \ain$), we can perform a second order expansion of the $S$-term by expanding the elliptic integral accordingly \citep{gradryz65}. We find
\begin{equation}
\lim_{\varpi \rightarrow 0} \; \tir S = -(1+s)-\uout^{-2}\frac{(s-1)\chi_{s-1}}{\chi_{s+1}},
\end{equation}
and then the ODE becomes
\begin{equation}
\tir \frac{d (\tilde{\psi}-1)}{d \tir}  - (1+s)(\tilde{\psi}-1) \approx - \frac{a_0^2 \sin^2 \theta}{\aout^2} \tir^2,
\label{eq:odeapproxu}
\end{equation}
whose solution is
\begin{equation}
\tilde{\psi}(\tir) \approx 1 + \frac{\chi_{s-1}}{\chi_{s+1}} \uout^{-2}.
\label{eq:psinear0}
\end{equation}
At second order, the potential in the inner domain ($r \ll \ain$) is quadratic with the cylindrical radius $R$ (while the gravitational acceleration is linear).

\section{The softening length at large relative separations}
\label{sec:ap:sl_asymptotics}

As in \cite{hp09}, we consider a vertical stratification of the form
\begin{equation}
\rho(z)= \rho_0 \left[1-\left(\frac{z}{h}\right)^{2q}\right]
\label{eq:vertprofile}
\end{equation}
for $|z| \le h$ (and $0$ elsewhere), where $\rho_0$ is the density at the disk midplane, $h$ the local semi-thickness (both a function of the radius $a$ in general), and $q \ge 1$ an integer. Homogeneous profiles correspond to $q \rightarrow \infty$, whereas the parabolic profile is obtained for $q=1$. We can compute the contribution of vertical stratification to the potential from the integral
\begin{eqnarray}
\chi 
     & = \chi_0 + \frac{1}{2q} \left( \chi_0 - \chi_q \right),
\label{eq:chi_mix}
\end{eqnarray}
where
\begin{equation}
\chi_q = (2q+1) \int_0^{1}{\left(\frac{z}{h}\right)^{2q} k \elik(k) d \frac{z}{h}}.
\label{eq:chiq}
\end{equation}
The approximations $k \approx 1$ and $\elik(k) \approx \ln 4/k'$, assumed in \cite{hp09}, is only valid for $a\approx R$. To derive the asymptotic behavior at large relative separation, it is sufficient to consider the case $k \rightarrow 0$, that is, at second-order
\begin{equation}
k \approx m \left[1 - \frac{1}{2} \left( \frac{h}{a+R}\right)^2 \right]
\end{equation}
as well as  $\elik(k) \approx \frac{\pi}{2}$. We then find
\begin{equation}
\chi_q = \frac{\pi}{2} m \left[1 - \frac{2q+1}{2(2q+3)} \left( \frac{h}{a+R}\right)^2 \right],
\label{eq:chiq2}
\end{equation}
and so
\begin{equation}
\chi =  \frac{\pi}{2} m  \left[1 - \frac{2q+1}{6(2q+3)} \left( \frac{h}{a+R}\right)^2 \right].
\end{equation}
The softening length is then deduced by equating $\chi = \ms  \frac{\pi}{2}$, where $\ms$ is the softened modulus:
\begin{equation}
\ms = \frac{2\sqrt{aR}}{\sqrt{(a+R)^2 + \lambda^2}}.
\label{eq:msoft}
\end{equation}
We then have
\begin{equation}
\frac{\lambda}{h} = \sqrt{\frac{2q+1}{2(2q+3)}}.
\end{equation}

\end{document}